# IMPLEMENTATION OF A STREAM CIPHER BASED ON BERNOULLI'S MAP

Ricardo Francisco Martinez-Gonzalez[1] and Jose Alejandro Diaz-Mendez[2]

[1]Electrics and Electronics Department, Veracruz Institute of Technology, Veracruz, Mexico.
[2]Electronics Department, National Institute of Astrophysics Optics and Electronics, Tonantzintla, Mexico.

## ABSTRACT

*A stream cipher was implemented on a FPGA. The keystream, for some authors the most important element, was developed using an algorithm based on Bernoulli's chaotic map. When dynamic systems are digitally implemented, a normal degradation appears and disturbs their behavior; for such reason, a mechanism was needed. The proposed mechanism gives a solution for degradation issue and its implementation is not complicated. Finally, the implemented cipher includes 8 stages and 2 pseudo-random number generators (PRNG), such cipher was tested using NIST testes. Once its designing stage, it was implemented using a developing FPGA board.*

## I. INTRODUCTION

Cryptography is the science of using mathematics for encrypt and decrypt data. The cryptography allows storing or transmitting important information, as it cannot be read for aliened entities [1]. Some of the most important cryptographic tools are ciphers, coders and water makers; however, among them, the ciphers are probably the most commonly used.

By the way the ciphers process data, they can be classified as: stream ciphers, this ciphers work over individual characters from plaintext at once, and block ciphers that take block by block from plaintext to cipher.

According to Kohel [2], the block ciphers are memoryless, due they use the very same function in order to cipher successive blocks; meanwhile, the stream ciphers must have memory, mostly because keystream is a function of initial value and system current state.

Strength in stream ciphers relies on keystream generating function that is the main reason to study and develop so many and various ones. One of used methods are chaotic functions, their functioning is based in certain behavior presents at some dynamic systems to naturally produce random sequences [3].

In table 1 from [4] is presented a similarity relationship between wanted features in cryptographic systems and natural features in chaotic ones. Among chaotic system features are its high sensibility to initial conditions, they have a pseudo-random behavior, and they are able to disperse data around their working space [5]. For such features, this class of function was chosen as keystream generating functions for present work.

## II. MATHEMATICAL DESCRIPTION FOR USED BERNOULLI'S MAP

The used function for generating keystream is a chaotic one. In the chaotic functions, one-dimensional maps are the simplest ones in order to generate chaotic sequences, and the selected one for this cryptographic system was Bernoulli's, mostly because this map has a simple digital implementation [7].

Tsuneda et al [8] presented a modification for Bernoulli's map; the modified map has next mathematical expression:

$$x_{i+1} = \begin{cases} 2\mu x_i + \frac{(1-\mu)}{2} & 0 \leq x_i < 0.5 \\ (2\mu x_i - 1) + \frac{(1-\mu)}{2} & 0.5 \leq x_i < 1 \end{cases} \quad \text{Eq.1}$$

Eq. 1 was originally presented to satisfy analog implementation, ergo to implement this expression using a binary-representation digital system is needed some manipulations of it.

After some mathematical manipulations, Eq. 2 is obtained; this expression is adequate for digitally implementing.

$$x_{i+1} = \begin{cases} 2\mu x_i + \frac{2^{bits}(1-\mu)}{2} & 0 \leq x_i < 2^{bits-1} \\ (2\mu x_i - 2^{bits}) + \frac{2^{bits}(1-\mu)}{2} & 2^{bits-1} \leq x_i < 2^{bits} \end{cases} \quad \text{Eq.2}$$

Sequences obtained from mathematical expression defined in Eq. 2 exhibit some non-randomness issues. The issues are a normal degradation caused by digital systems flooring [9]. Some authors had proposed some partial solutions; one of them is implementing several stages [10] in order to increase its randomness.

## III. BIFURCATION DIAGRAMS

Baker and Gollup [11] define bifurcation as system behavior determination at control parameters variations, its most useful representation takes place when in a system; its control parameters are consequently varied, this representation is known as bifurcation diagram [12].

In a bifurcation diagram, horizontal axis represents µ parameter and vertical axis represents higher iterations $f_\mu^n(x_0)$ for a specific initial point $x_0$; in consequence, specified diagram depicts $x_0$ orbit behavior. Figure 1 depicts obtained bifurcation diagram following Elaydi's procedure [13] for Bernoulli's map.

The form of bifurcation diagram is representative of implemented function. As it was previously discussed, discretely implemented dynamic systems have a degraded behavior; thus, a solution is required. The solution jumped out after seeing bifurcation diagram for 4 symmetric portions of obtained sequences from the original system.

In shown diagrams at Figure 2, first and most significant portion of bifurcation diagram, the most significant, is quite similar as the presented one in Figure 1. In the other hand, the bifurcation diagrams for other the three portions of the original sequence present very different diagrams.

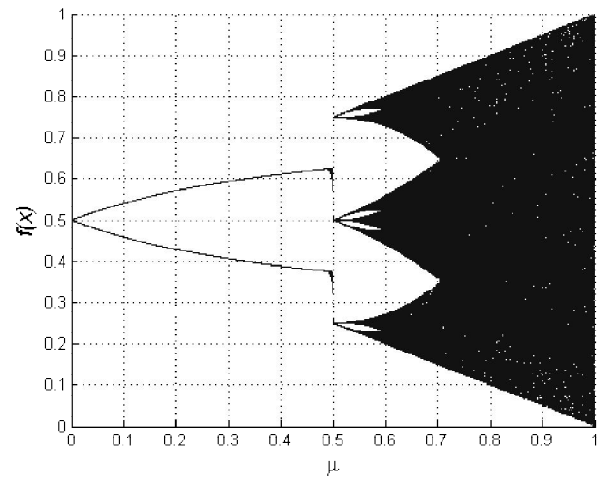

Figure 1. Bifurcation diagram for Bernoulli's map.

The orbits in second, third and fourth portion of bifurcation diagrams are along entire space, with such behavior the to-be-ciphered data might be equally dispersed along entire space as well.

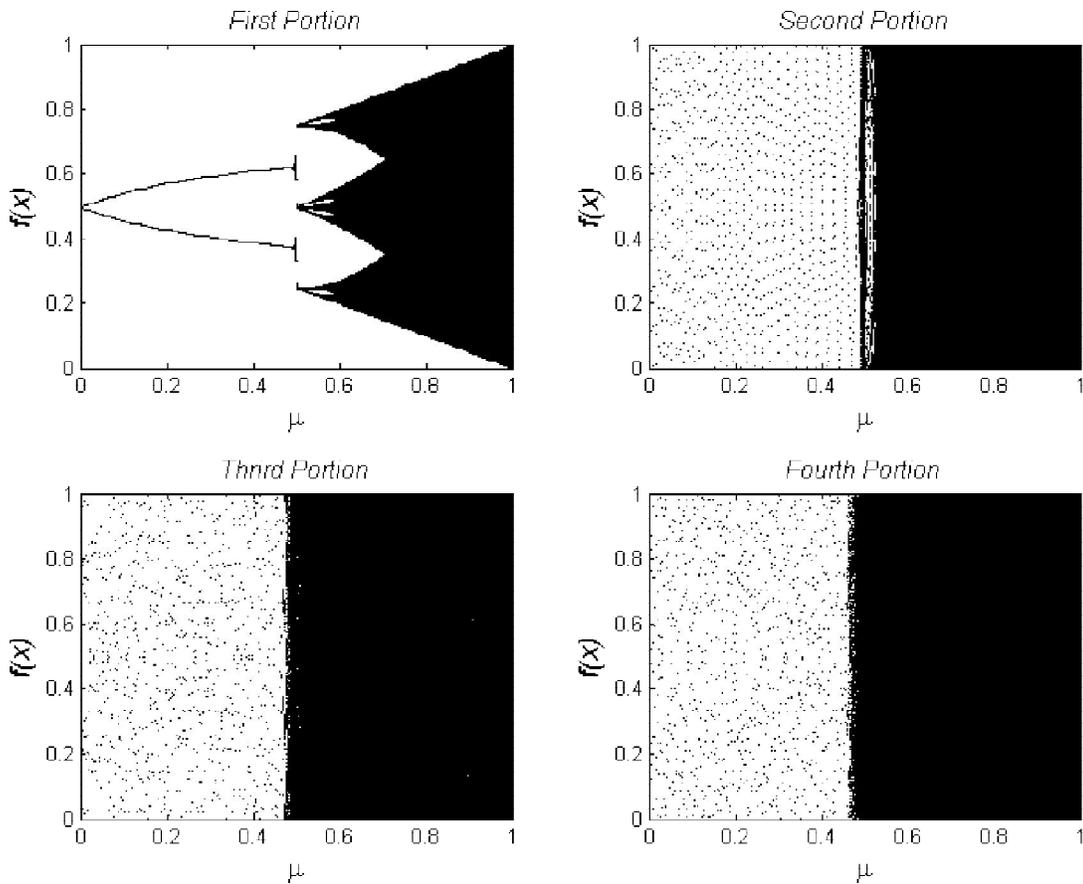

Figure 2. Bifurcation diagrams for sections 2, 3 and 4 from Bernoulli's map original sequence.

Prior finding caused to design a structure that separates each original 32-bit sequence into four 8-bit sequences. The keystream used by the cryptosystem is result of a XOR procedure with eight 8-bit sequences, because the cryptosystem contents 2 PRNG.

## IV. 8-BIT SEQUENCES GENERATING

In order to obtain 8-bit sequences, firstly is necessary to separate each original 32-bit sequence into 16-bit sequences; the mathematical expressions that carry the task out, is defined in Eq. 3 and Eq. 4.

$$f(x)_{first\ part} = floor\left(\frac{f(x)}{2^{bits/N}}\right) \quad \text{Eq. 3}$$

$$f(x)_{second\ part} = mod(f(x), 2^{\frac{bits}{N}}) \quad \text{Eq. 4}$$

Using Eq. 3 is obtained the most significant half of f(x). The procedure is based in the fact that in binary representation, right-shifting is done dividing by 2n, where n represents desired places to shift. The second half is obtained by same prior principle; even though in this occasion, division residue is the required part; mathematical description for such task appears in Eq. 4.

Once both 16-bit sequences were obtained, each sequence is separated using again Eq. 3 and Eq. 4 in order to obtain four 8-bit sequences. In addition, to increase cipher complexity, two 32-bit sequence generators are used, generating eight 8-bit sequences in total, with them the cryptosystem reaches an acceptable security level.

## V. VHDL IMPLEMENTATION FOR PROPOSED CIPHER

For FPGA implementation, an Altera's Cyclone IV EP4C22F17C6N was selected. The cryptosystem was firstly simulated in ModelSIM, and matching results using Matlab. Because of cipher features, its design is separated in two. The first part const of 2 PRNG and a XOR-gate array is second part. Both PRNG need an initial value to start generating their sequences. After initial value is introduced, the PRNG feedback loop needs to be closed; on account of it, a mechanism was figured out.

The mechanism utilizes a D-type flip-flop, its Q-output is definitively set to logical-one after PRNG's first operation. The feedback loop is controlled by a MUX, whereas the MUX is controlled by the Q-output. MUX afterward, its output is multiplied by two, with a throwing 33-bit overflow condition, then the multiplier result is multiplied by feedback factor (μ), as well as initial value, μ is a control parameter.

Output of feedback factor multiplier has 40 bits; the least significant 8 bits are eliminated and the remained ones go to an adder that adds it with generalization factor defined in Eq. 5.

$$\frac{2^{bits}(1-\mu)}{2} \quad \text{Eq. 5}$$

Finally, the output adder is saved by a parallel input/parallel-output register; furthermore, the register helps controlling out PRNG flow. Figure 3 depicts block diagram for the described PRNG.

International Journal of Computer Science & Information Technology (IJCSIT) Vol 6, No 6, December 2014

The sequences obtained from the PRNG got separated in 4 equal-length sequences; using 2 PRNG, eight 8-bit sequences are obtained.

Next part to describe is a XOR-gate array, which is in charge to mix data up. The most significant bit from each sequence passes through XOR gates in order to obtain keystream most significant bit; the next significant bits pass through another XOR-gate group, and so on until the least significant bits.

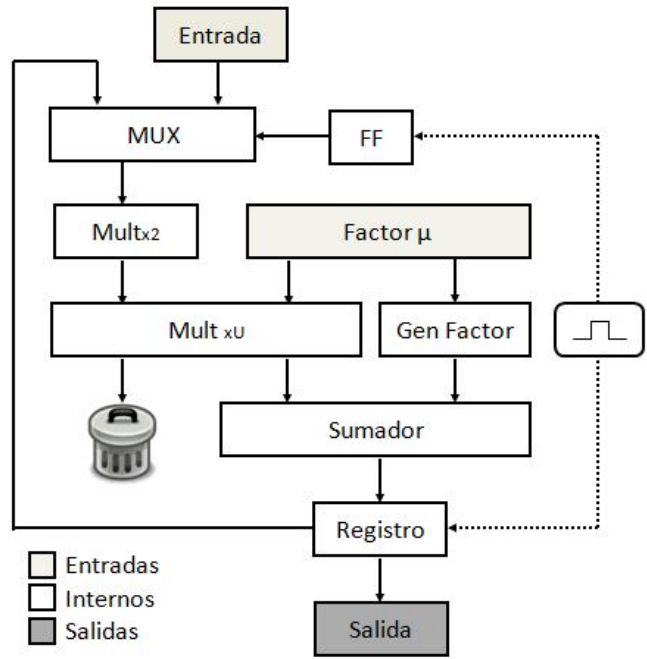

Figure 3. PRNG block diagram.

The whole array consists of 56 XOR-gates, 7 for each keystream bit, the array may be observed on Figure 4. The XOR-gate array output is the keystream used by proposed stream cipher.

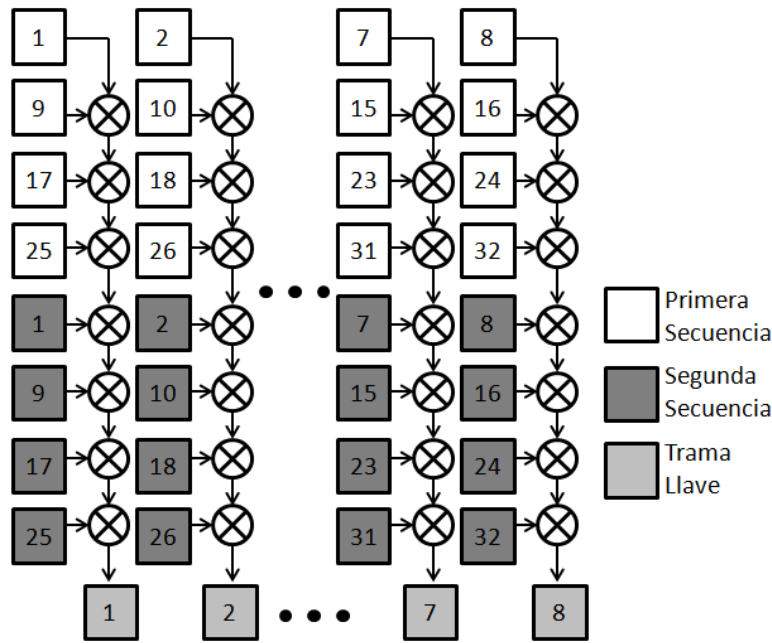

Figure 4 Disposition of XOR-gates in-charge of obtaining keystream

## VI. SIMULATION RESULTS

ModelSIM was used to realize simulations for the proposed stream cipher, and a Matlab script was made to verify simulation result. Figure 5 shown results for PRNG ModelSIM simulation.

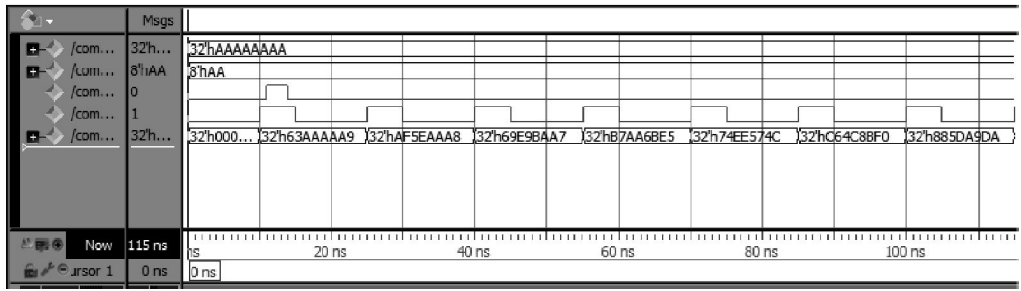

Figure 5. PRNG simulation using 2863311530 as initial value and 0.6640625 as µ factor

In figure 5, "32hAAAAAAAA" correspond to initial value, which starts PRNG up; the next value "8hAA" is µ factor. Signal 4, from top to bottom, is clock system. Signal 3 is a pulse for the closing feedback loop flip-flop. The last signal is output PRNG, only few data were chosen due data presentation; nevertheless, data is exactly same as obtained one by Matlab; moreover, high sensibility to initial conditions in chaotic systems [14] inspires reliability between obtained Matlab data and FPGA one, even though only few data is presented.

Once PRNG function has been verified by simulation, next simulation is for separating sequences mechanism, simulation result is shown at Figure 6. In Figure 6, the first two signals, from top to bottom, are initial values; "32hAAAAAAAA" for first PRNG and "32hBBBBBBBB" for second one.

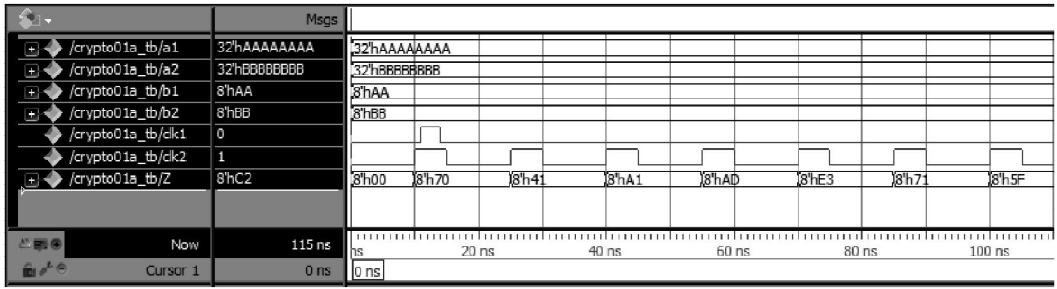

Figure 6. Simulation of keystream generating mechanism for proposed cipher.

The next pair of signals is µ factors, first PRNG receives "8hAA" and second one receives "8hBB". Signal 6 is system clock, and signal 5 is in charge of controlling feedback loop closure. Finally, signal 7 is output stream cipher; in other words, the keystream for the proposed cryptosystem.

Results obtained in FPGA were completely matched with simulation ones. Besides VHDL codes just needed some minor modifications, FPGA implementation was successful and untroubled.

## VII. NIST STS TESTING

The NIST Statistical Testing Suite (STS) is based on determining whether or not a specific sequence of zeros and ones are random [15].

The NIST STS was developed to test the randomness of binary sequences produced either hardware of software based cryptographic random or pseudorandom number generators [16]. In NIST STS, testing results is P-value. It may have values between 0 and 1and the bigger P-value is, the better pseudorandom property the tested sequence has [17]

The proposed implementation was tested in order to know how good it is. A random sequence was elaborated using $\mu_1$=0.75 and $\mu_2$=0.8, and $1.2885 \times 10^9$ as initial value for first generator and $8.5899 \times 10^8$ for the second one. The random sequence had 4 million bits, and the obtained results are presented in Table 1.

Table 1. NIST STS testing results.

| Test | Obtained P-value | Status |
| --- | --- | --- |
| Frequency test | 0.988 | OK |
| Block frequency test | 0.986 | OK |
| Run test | 0.987 | OK |
| Cumulative sum test (forward) | 0.988 | OK |
| Cumulative sum test (reverse) | 0.988 | OK |
| FFT test | 0.998 | OK |

With Table 1 results, it is plausible to ensure that the proposed cipher works with a good chaotic behavior, and in consequence, good cryptographic features.

International Journal of Computer Science & Information Technology (IJCSIT) Vol 6, No 6, December 2014

## VIII. CONCLUSIONS

Current developing started trying to solve a chaotic function instauration in a stream cipher design. The problem was solved adequately, but another problem showed up, a normal degradation, it affects digitally-implemented dynamic systems. The new problem was solved implementing a dividing sequence mechanism; and sequences were obtained by two PRNG. The cryptosystem was verified using a NIST STS.

Using Matlab helps out to verify simulation results, the method makes easier to manage so many and large data. In conclusion, cryptosystem implementation in a FPGA is easy once a mathematical analysis and an adequate VHDL coding simulation were made.

## Authors

**Ricardo F. Martinez-Gonzalez, PhD.**

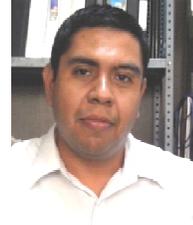

He was born on Veracruz, Mexico. He received his Bachelor of sciences degree from Veracruz Institute of Technology. Next on, He studied his M.Sc. degree at National Institute for Astrophysics, Optics and Electronics, in this place He started to get involved with Chaos issues. For obtaining his degree, He wrote how design pseudo random number generators using one-dimensional chaotic maps. Once He obtained the degree, He got into Veracruz University for a five month period; nevertheless, He returned to continue his research in Chaos subject, and get it to the next level, designing ciphers. He took several courses about digital communications and communication protocols; such courses help him out to finish his research and finally obtain his PhD degree. At this point, He took the choice of return to his basics; since then, He works at Veracruz Institute of Technology, where he has been enrolled into a research crew.

**J. Alejandro Diaz-Mendez, PhD.**

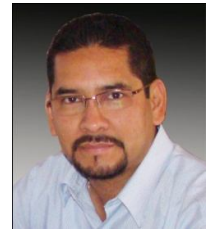

Dr. J. Alejandro Díaz-Méndez is a Full Researcher in the Electronics Department at INAOE. He received his BSc degree from Universidad Veracruzana, México; followed by M.Sc and Ph.D. degrees from Instituto Nacional de Astrofísica, Óptica y Electrónica (INAOE), México, in 1995 and 1999 respectively. In 1999 he was appointed as professor-Researcher at Instituto Politécnico Nacional, México. He has authored 30 Journal papers and around 80 conference papers. He has been a member of technical committees in national and international conferences.  He is an IEEE Senior member and a member of national researcher´s system of México.